# Identifying Potential Inlets of Man in the Artificial Intelligence Development Process

Man and Antiblackness in AI Development

This work identifies ways in which the foundations of AI development routinely lead to the creation of antiblack technologies

Deja Workman

College of Information Sciences and Technology, Pennsylvania State University, dejaworkman@psu.edu

Christopher L. Dancy

Department of Industrial and Manufacturing Engineering & Computer Science and Engineering, Pennsylvania State University, cdancy@psu.edu

In this paper we hope to identify how the typical or standard artificial intelligence development process encourages or facilitates the creation of racialized technologies. We begin by understanding Sylvia Wynter's definition of the biocentric Man genre and its exclusion of Blackness from humanness. We follow this with outlining what we consider to be the typical steps for developing an AI-based technology, which we've broken down into 6 stages: identifying a problem, development process and management tool selection, dataset development and data processing, model development, deployment and risk assessment, and integration and monitoring. The goal of this paper is to better understand how Wynter's biocentric Man is being represented and reinforced by the technologies we are producing in the AI lifecycle and by the lifecycle itself; we hope to identify ways in which the distinction of Blackness from the 'ideal' human leads to perpetual punishment at the hands of these technologies. By deconstructing this development process, we can potentially identify ways in which humans in general have not been prioritized and how those affects are disproportionately affecting marginalized people. We hope to offer solutions that will encourage changes in the AI development cycle.

**ACM Reference Format:**
First Author's Name, Initials, and Last Name, Second Author's Name, Initials, and Last Name, and Third Author's Name, Initials, and Last Name. 2018. The Title of the Paper: ACM Conference Proceedings Manuscript Submission Template: This is the subtitle of the paper, this document both explains and embodies the submission format for authors using Word. In Woodstock '18: ACM Symposium on Neural Gaze Detection, June 03–05, 2018, Woodstock, NY. ACM, New York, NY, USA, 10 pages. NOTE: This block will be automatically generated when manuscripts are processed after acceptance.

# 1 INTRODUCTION

Sylvia Wynter's concept of the biocentric Man genre of *the human* describes the flattening of all of the human existence into the entity of a single, pseudo-imaginary person, which we will refer to simply as "Man" heretofore for simplicity [1]. The biocentric Man is a concept that is ill-representative of the whole human population because the conditions that created the notion of a biocentric Man exclude those who, by its definition, are *other* [than human]; these interlocking systems and conditions of white-supremacy, heteropatriarchy, and capitalism enforce boundaries [2,3] and discriminates based on race, gender, sexuality, and class to dehumanize and other people who do not neatly fit into what has been defined as the ideal human. Still, this is a concept that the world's everyday activities operate around, and the sociotechnical world is no exception. Modern service to the biocentric Man is often unintentional or undetected and can come in many forms, but all of these forms are exclusionary. The prevalence of the biocentric Man is especially concerning because, firstly, we've not consistently prioritized human-centered design, meaning that at times we value the impact that our technology has on other factors (i.e. productivity or accuracy) more than we value or consider the impact that our technology will have on people, whether that be the user of the technology or those who are tangentially affected by it. Secondly, because of our dependence on the biocentric Man to stand in as a substitute for humanity, when we do take the additional steps needed to recenter the human, we end up recentering on the biocentric Man genre of the human instead.

So with the de-prioritization of the human in development processes, as the presence of artificial intelligence-based systems continues to grow in our society, so does their potential to be harmful. We have seen examples of these harms more frequently with this increasing ubiquity. One example of these realized harms was identified when it was proven that an AI-based diagnostic tool was underdiagnosing under-served patients—that is, this technology was more likely to report that Black people, women, and people from lower socioeconomic statuses were healthy when they in fact were ill [4]. People in these categories were more likely to not receive any treatment for their condition because this diagnostic tool was not adequately prepared to diagnose people who were not white men. Another example would be in the use of surveillance technologies by government entities, especially police. Facial recognition technology has produced a global concern, but just among the Black community in America, there are multiple aspects of this technology to consider. Facial recognition technology is more likely to misidentify people with darker skin tones; these technologies produce a likeness score between faces to match potential subjects, however it is up to a human employee to make a final call on whether or not a likeness score is high enough to consider the person a match. This introduces all of the bias of humans and all of the authority of technology to create an extremely dangerous environment for people left in the hands of law enforcement [5,6,7]. Not only that, but the creation of this technology is blind to the history of Black communities being policed, surveilled, and all too often killed at the hands of the very same people who we've now equipped with AI systems [8]. Outside of the United States, we've seen the same types of technology serve as spyware used to shut down protests, arrest social activists and quiet civil unrest so that unjust governments can continue calmy oppressing their own citizens [9]. The existence of these AI systems, especially in the hands of institutions that work to maintain Man poses a threat to the safety and privacy of Black people.

The numerosity of these problematic consequences are more than the result of periodic mistakes that happen more frequently due to the rapid expansion of artificial intelligence, this is the result of the fact that the development processes we use today were made to serve the biocentric Man. Designing technology, or particularly AI as in our case, in a way that invites, appeals, and attends to the experience of people—regardless of their social profile—is incredibly complex. The way forward is to recenter these technologies on humans as a whole and encourage a foundational shift that insists upon the careful consideration of diverse perspectives in the context of overlapping systems of oppression that result in the biocentric Man. In this paper, we will define and dissect our interpretation of a formalized, generic artificial intelligence development process with the intention of identifying common practices that remain more susceptible to producing malignant outcomes, particularly antiblackness [10].

# 2 UNDERSTANDING THE BIOCENTRIC MAN GENRE OF THE HUMAN

Grasping the idea of the biocentric Man genre is an important part of understanding the issue of antiblackness in the AI development process because the same people being othered by the definition of the biocentric Man are the same people being disproportionately harmed by the AI-based systems that eventually end of having unintended consequences. Broadly, the biocentric Man is characterized by the western White, middle-class, cisgender man, it is no one in particular but rather the notion of one—as indicated by the term *genre*—a momentary frame or form, culture or kind of human. And it is such due to the history of wealth, colonialism, and position of power that is tied to the western White man; the same power that allotted them the ability to define humanness, which was used to deselect variations from the *norm*. Accordingly, society, sociotechnical systems, and (particularly) AI systems functionally accommodate this biocentric man genre of the human, with Black people, and many more, removed from the definition of the biocentric Man, society is (and thus sociotechnical systems are) not built with them in mind. Technology is not built with them in mind. Wynter says that our way forward is to reconceptualize our representation of the human, and we argue that by reinviting previously excluded genres of human back into the structures that we continue to build, we can begin to achieve that [1].

## 3 THE AI DEVELOPMENT PROCESS

As we stated previously, we'll begin by establishing a tangible process that we believe to be broadly reflective of current AI development processes. Having a single model such as this will be beneficial for an organized deconstruction. In this deconstruction we will more specifically highlight ways in which these steps play to the biocentric Man. We've developed this representation of the process by reviewing available processes from places such as Microsoft, the CRISP-DM Model, the field of SE4AI, several DevOps and MLOps guides, and many more. It is important to recognize that not all institutions (i.e. universities, corporations, etc.) use the same development process, and there may even be multiple development processes being used within the same institution. However, we still managed to identify some common themes and shared steps that took place across several different institutions. We will not individually delve into every single sub-step that we've decided to include in our development model, but we will choose which ones seem best-fitting according to the intentions of the paper. We highlighted these steps in Figure 1, where we break down the AI development process into six broader stages: identify a problem, selection of development processes and resource management tools, dataset development and data processing, model development, deployment and risk assessment, and integration and monitoring [11, 12, 13].

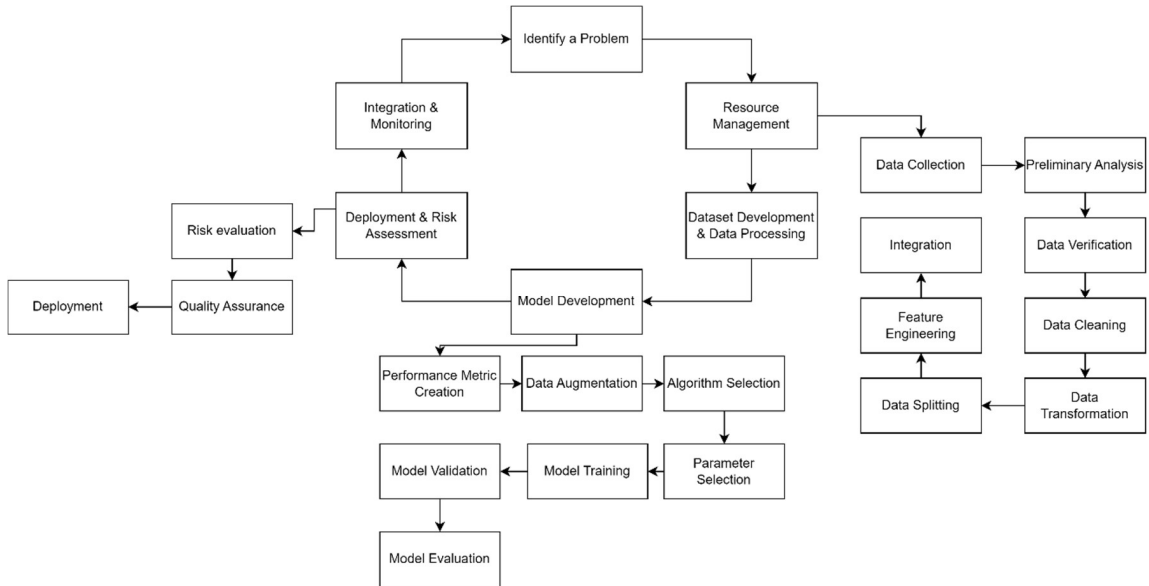

Figure 1: The stages of our model AI Development Process.

### 3.1 Identifying Problems and AI-Based Solutions

The identification of a problem that we hope to resolve can be thought of as the "why" while the resource management stage can be thought of as the "how". The label of "problem" is misleading in a way, as there does not necessarily need to be a problem in order for an institution to begin developing AI—this has always been true but in some ways it has become more obvious as our technological capabilities advance, and it is an observation that both precedes and exceeds artificial intelligence. In fact, some would argue that there has been a scientific shift past creating technology for the benefit of the people, and onto creating technology for the benefit of organizations and companies, showing that there are times in which profits are taken more seriously than these disproportionate impacts. This thought has especially been revisited as OpenAI made headlines in 2022. It has long since been discovered that OpenAI subjected poorly paid workers, who were very often people of color, to traumatizing content in order to create the GPT-4 dataset, which stores all of the data used by Chat-GPT [14]. For around $2 an hour, employees sifted through copious amount of text which often included explicit content pertaining to gore, violence, and sexual abuse and thus some employees have come out stating that they now suffer from workplace induced PTSD. This was a common enough narrative that counselors were additionally hired for the employees, but that simply was not enough. The broadcasting of this knowledge has not appeared to curb the popularity of Chat-GPT, and the unethicality of how it came to exist is rarely ever mentioned. The fact that OpenAI has not faced any real form of repercussion for its actions is a direct example of how we've generally disregarded the harm to poor people of color caused by an AI-based technology—we can see this lack of accountability in the repeated failure to mention the labor behind this tool as time passes and in the profits that OpenAI continues to have. This is the main point of what we want to highlight in this stage, we routinely find that marginalized people are disproportionately impacted in the decision-making process that leads to the generation of new technologies because we don't consistently consider the motivations, contexts, or broader impacts of the work.

### 3.2 Selecting Development Processes and Managing Resources

Resource management is an unconfined stage, because it can affect the order and iteration of the remainder of the cycle. Our process is configured not unlike that of a waterfall development process for the sake of having a simple process to deconstruct, but modernly we tend to more often see processes that are less sequential. This will change how and when and how often each of the stages of the project is completed, but in one way or another, each stage will remain present. This is especially important because more iteration and evolutionary approaches may be able to produce results that differ from those of sequential development models [15]. The term resource management refers to any and all approaches or tools to how the AI development process will be completed—ranging from philosophies to frameworks—despite the fact these do not individually belong in the same category outside of our particular analysis.

### 3.3 Dataset Development and Data Processing

The dataset development process can vary based on the project needs and the availability of the data, but it is an extremely important stage that has many factors nonetheless. Thus, we've broken this stage into an additional 8 sub steps: data collection, preliminary analysis, data verification, data cleaning, data transformation, data splitting, and integration. The data collection sub step includes the initial sourcing and gathering of data that will be needed for the construction of the dataset. When we are considering how to decenter the biocentric Man from this process, data collection and data cleaning become particularly critical. The process of data collection is naturally extractive and all things extractive have the potential to be exploitive. To recenter real people, we should return what we take from these communities; the people who we've taken this data from should benefit from what we make of it. Data collection also has a tendency to instill data biases and advocate for the Man due to poor sourcing techniques. When dataset developers are not educated on how the source of data might impact the characteristics of the data itself, we risk reifying systems both in this process and substantially reproducing them in the model development stage.

Data cleaning can easily become exclusive. The terms "unclean" or "messy" data are especially risky because it allows those who are cleaning the data to subconsciously infer for themselves what, or better yet who, *clean* data looks like [16]. Defining an outlier, deciding to remove or keep an outlier, and justifying what one needs to do to or change about data are all so such subjective yet high stake decisions. It is a decision of whether or not someone, or a genre of people, get to be represented in technology that may come back to affect them. It is, yet again, the power of deciding who will be allowed here as themselves, and who will be othered.

### 3.4 Model Development

The model development stage applies machine learning to the dataset created previously to identify patterns and insights within it, we've broken this stage up into the following 7 sub steps: performance metric creation, model training, data augmentation, algorithm selection, parameter selection, model validation, model evaluation. Performance metrics are established both as something for the model to optimize for as well as something to judge when comparing the performance of multiple models. Optimization and the biocentric Man genre are closely related because in choosing what to consider or develop as a performance metric, model developers may inadvertently be deciding on whom to optimize for. This often comes back to who is represented in the dataset used to train the data, which is, of course, highly related to the model training stage. Performance metrics can also be used as indicators of what inputs do or don't fit properly into the model.

Choosing a performance metric is often regarded as a straightforward process but it can still have unexpected consequences, which is sometimes referred to as the cobra effect [13]. To put this back into the perspective of the biocentric Man, a model that is able to achieve a high level of accuracy by accurately evaluating 90% of a dataset despite continuously inaccurately evaluating the remaining 10% is still often seen as accurate according to a performance metric creation that looks at the totality of accuracy—which many of them do. The performance metric, however, often would not be developed to account for the fact that everyone in the 10% has the same or similar relations to existing systems of oppression (and, thus, the biocentric Man) that results in the model's . The performance metric on its own as they are currently and typically being created does not guarantee thorough feedback on who the 10% is, only that the model has a 90% accuracy—the performance metric reports as though the accuracy is random, but it becomes clear that this is not the case when developers eventually realize that the 10% had a racial differentiation from the remaining 90%. An additional layer to this problem is introduced when considering the comparison aspect of the performance metric. As they are interpreted now, in a comparison between two models which each received a dataset a performance metric that only considers total or summarized accuracy can be misleading. If the first model correctly identifies 100% of the larger subset but 0% of the smaller subset, and the second model correctly identifies 90% of the larger subset as well as 50% of the smaller subset, then the first model would still be considered more accurate despite the fact that it could not accurately assess anyone in the smaller subset. This is an extremely simplified version of what's happening in the real model development process for the sake of explanation, but the concerns are the same. While this stage is highly regarded as being straight-forward, there is enough evidence that perhaps it should not be regarded as such.

Another notable sub-step in the model development stage is data augmentation, which employs statistical methods to produce additional data to be used for model training. This can take place in the form of bootstrapping, which we will focus on here. While data augmentation is a necessary step in model development, and while it may be neither reasonable nor possible to bypass this step, it remains important to at least acknowledge the residuals of the bootstrapping process to further contemplate how we might be able to counter or rectify them. Bootstrapping replicates what it sees in the initial model training dataset and when this initial dataset replicates existing white supremacist relations and racist content then the environment produced by this bootstrapping procedure also turns out to be racist. This is no surprise, but regardless of expectation this is still incredibly harmful, especially considering that the output will be racist at a greater magnitude due to the numerosity of these replicants.

### 3.5 Concluding the Development Process

As we near the later stages of the development cycle, it may or may not become less clear how the Man is being centered in the development process. Still, each of the steps present—including those steps that are responsible not for the creation of the model but rather for the movement of the model into an accessible environment where users will interact with it—plays a role in facilitating the rest of the process. The deployment and risk assessment stage addresses the risks of the evaluated model then moves the chosen model to the production environment. Risk Assessment varies by strategy, but most procedures share similar principles while differing in iteration and concurrency. The sub steps of this stage are environment adaptation, risk evaluation, quality assurance, and deployment. The extent of environment adaptation is dependent on the cooperation between the development and production environments, but in any case, this sub step is where that form of transformation takes place. This type of transition would also potentially be risky to the validation process if the change was not prepared for during the model development stage. In considering the preparation for these transformations, it is also important to consider the performance factors of the adaptation. These factors may include latency depending on languages used, power, server advancements, and data access. All of the mentioned factors can result in a reified biocentric Man as acceptable environments and data access are connected to what is seen as important for a production-ready AI system. Power and server advancement choices can have disproportional effects on communities composed of folks not considered a part of the biocentric Man and data that is accessed is too often only representative of Man or the relations to antiblackness seen in certain, easily accessible, online environments [17]. Furthermore, data access can have a more obvious impact on the model, since it introduces the decision of data updates, decisions of which are impacted by antiblackness and Man. Models can either be directly connected to a stagnant form of data or a database that can be updated as needed, the former of which poses risk to the relevancy of the model after a period of time.

This brings us to consider the remaining risks associated with the movement of the model into the production environment. These risks most often appear in the form of bugs, poor training data, high error rates, dissimilar training and production environments, and misuse of the model. Quality assurance throughout the other stages of the development process can reduce both the intensity of the quality assurance at this point and should be done at more points than one. Additional stats surrounding accuracy and efficiency of the model are collected and developers can use this information to decide whether or not the model is ready to be deployed. This same information can be used at a later point in time to compare the current model with a future, newer version of the model [18].

In our representation of integration and monitoring, the model has already been deployed to the production environment. This stage regards the maintenance and upkeep of the model throughout its life and consists of two sub-steps: establishing feedback loops and model updates. Feedback loops typically include first forming a logging system for the feedback loop to store information in, this information will include some of the same stats that we've seen up to this point. That data is then used in review to reevaluate the model to see if it needs to be retrained based on a decrease in accuracy rates, called drift. When creating new models to replace a model that has drifted, model developers will typically employ either shadow testing or A/B testing to compare the different models and update the production environment with the highest preforming new model. The feedback loop and model update sub-steps are quite intertwined, as the actual model update process itself is resemblant of the deployment process, but the contents of the update are dependent upon the feedback generated from the feedback loop.

## 4 RACIALIZED ARTIFACTS AND CREATING SOLUTIONS

The majority of this paper is meant to analyze the potential sources of antiblackness in the AI development process, but this analysis needs to be applicable in order for us to see any real movement away from the default catering to the biocentric Man. It's likely that this change will not arise from any one approach or framework, but rather from a foundational reconstruction that will more strongly encourage the consideration of the results of our decision making and who those results will impact. There is also great merit to understanding each of the concerns that we presented and developing artifacts to address them individually, and there is progress being made among many of these inlets. While the initial stages of our research are more dedicated to understanding the structure that we're deconstructing and seeing where the cracks are in that formation, we have also begun to discuss possible future solutions.

Beginning with what was discussed in the problem identification stage, we found that these inlets were majorly rooted in historical discrimination. The long-term effects of systematic racism that Black Americans have been the victim of for decades on end has caused us to be inherently more vulnerable to exploitation, as we've been socially economically and politically marketed as such, and this fact does not stop at the border of technology. The disproportionate consequences inflicted upon Black Americans is not treated with the same concern or urgency of that of White Americans and thus these consequences continue to multiply. We continue to produce harmful technologies so long as it is only at the expense of Black bodies. We continue to produce technologies without considering how they will affect Black communities. Moving towards a solution means encouraging (or mandating and demanding) that people with the power to bring these technologies into existence become aware of the social implications of their work. We need to reclaim creation as a term reserved for creations that do not cause additional destruction and with that, we need to see and understand the fact that not every invention or application is a form of technological progression; progress is defined by movement forward and I will argue that progress for some at the regression of others is not really progress at all. Which brings us to consider action in the face of regression. AI ethics principles often highlight data and algorithmic bias as high-risk stages, and they are correct in doing so, but we would like to draw attention to the fact that these inlets are found as early on as the decision to begin the development process. Taking reforming actions at this stage is unpopular but we need to become comfortable asking why some things are being made at all.

## 5 CONCLUSIONS

Understanding how we can continue to remove the biocentric Man from the AI development cycle is a task that requires thorough knowledge of the development process and a contextual understanding of the history of the Black community in America. We will continue to explore this connection and think more specifically about certain parts of the model development process. In future work we plan on shifting more towards what the individual solutions to these problems might be and creating artifacts that support our goal of recentering real people in the AI development cycle.

# REFERENCES

[1] McKittrick, K. (2015). *Sylvia Wynter on being human as praxis*. Duke University Press.

[2] Parker, E. (2021). Elemental difference and the climate of the body. Oxford University Press.

[3] Browne, S. (2015). Dark matters: On the surveillance of blackness. Duke University Press.

[4] Seyyed-Kalantari, L., Zhang, H., McDermott, M. B., Chen, I. Y., & Ghassemi, M. (2021). Underdiagnosis bias of artificial intelligence algorithms applied to chest radiographs in under-served patient populations. *Nature Medicine*, *27*(12), 2176–2182. https://doi.org/10.1038/s41591-021-01595-0

[5] Najibi, A. (2020, October 26). *Racial discrimination in face recognition technology*. Science in the News. https://sitn.hms.harvard.edu/flash/2020/racial-discrimination-in-face-recognition-technology/

[6] Mora, A. R. (2020). Shades in gender: Visualizing gender diversity through color palettes. *Communication Teacher*, *35*(1), 43–48. https://doi.org/10.1080/17404622.2020.1797134

[7] *Project Green Light Map*. City of Detroit. (n.d.). https://detroitmi.gov/webapp/project-green-light-map

[8] Peeples, L. (2020). What the data say about police brutality and racial bias — and which reforms might work. *Nature*, *583*(7814), 22–24. https://doi.org/10.1038/d41586-020-01846-z

[9] https://www.washingtonpost.com/world/middle_east/israel-palestinians-surveillance-facial-recognition/2021/11/05/3787bf42-26b2-11ec-8739-5cb6aba30a30_story.html

[10] Dwoskin, E. (2021, November 8). *Israel escalates surveillance of Palestinians with facial recognition program in West Bank*. The Washington Post. https://www.washingtonpost.com/world/middle_east/israel-palestinians-surveillance-facial-recognition/2021/11/05/3787bf42-26b2-11ec-8739-5cb6aba30a30_story.html

[11] De Silva, D., & Alahakoon, D. (2022). An artificial intelligence life cycle: From conception to production. *Patterns*, *3*(6), 100489. https://doi.org/10.1016/j.patter.2022.100489

[12] Amershi, S., Begel, A., Bird, C., DeLine, R., Gall, H., Kamar, E., Nagappan, N., Nushi, B., & Zimmermann, T. (2019). Software engineering for machine learning: A case study. *2019 IEEE*. https://doi.org/10.1109/icse-seip.2019.00042

[13] Lefevre, L. D.-S. K. (2020). *Introducing MLOps*. O'Reilly Media, Inc.

[14] Perrigo, B. (2023, January 18). *OpenAI used Kenyan workers on less than $2 per hour: Exclusive*. Time. https://time.com/6247678/openai-chatgpt-kenya-workers/

[15] Pew, R., and Mavor, A. (2007). Human-System Integration in the System Development Process. https://doi.org/10.17226/11893.

[16] Rawson, K., & Muñoz, T. (2019). Against Cleaning. *Debates in the Digital Humanities 2019*, 279–292. https://doi.org/10.5749/j.ctvg251hk.26

[17] Bender, E. M., Gebru, T., McMillan-Major, A., & Shmitchell, S. (2021). On the dangers of stochastic parrots. *Proceedings of the 2021 ACM Conference on Fairness, Accountability, and Transparency*. https://doi.org/10.1145/3442188.3445922

[18] Mateescu, A., & Elish, M. C. (2019, January 30). *Ai in context*. Data & Society. https://datasociety.net/library/ai-in-context/